\documentclass[sigconf]{acmart}
\AtBeginDocument{%
  }

\copyrightyear{2025}
\acmYear{2025}
\setcopyright{acmlicensed}\acmConference[KDD '25]{Proceedings of the 31st ACM SIGKDD Conference on Knowledge Discovery and Data Mining V.2}{August 3--7, 2025}{Toronto, ON, Canada}
\acmBooktitle{Proceedings of the 31st ACM SIGKDD Conference on Knowledge Discovery and Data Mining V.2 (KDD '25), August 3--7, 2025, Toronto, ON, Canada}
\acmDOI{10.1145/3711896.3737117}
\acmISBN{979-8-4007-1454-2/2025/08}

\usepackage{subcaption}
\usepackage{multirow}
\usepackage{array}
\usepackage{balance}
\usepackage{multicol}
\begin{document}

\title{Scaling Transformers for Discriminative Recommendation via Generative Pretraining}

\author{Chunqi Wang}
\authornote{Corresponding author.}
\affiliation{
    \institution{Alibaba International Digital Commercial Group}
    \country{China}
    \city{Hangzhou}
}
\email{shiyuan.wcq@alibaba-inc.com}

\author{Bingchao Wu}
\affiliation{
    \institution{Alibaba International Digital Commercial Group}
    \country{China}
    \city{Hangzhou}
}
\email{wubingchao.wbc@alibaba-inc.com}

\author{Zheng Chen}
\affiliation{
    \institution{Alibaba International Digital Commercial Group}
    \country{China}
    \city{Hangzhou}
}
\email{angxiao.cz@alibaba-inc.com}

\author{Lei Shen}
\affiliation{
    \institution{Alibaba International Digital Commercial Group}
    \country{China}
    \city{Hangzhou}
}
\email{kenny.sl@alibaba-inc.com}

\author{Bing Wang}
\authornotemark[1]
\affiliation{
    \institution{Alibaba International Digital Commercial Group}
    \country{China}
    \city{Hangzhou}
}
\email{lingfeng.wb@alibaba-inc.com}

\author{Xiaoyi Zeng}
\affiliation{
    \institution{Alibaba International Digital Commercial Group}
    \country{China}
    \city{Hangzhou}
}
\email{yuanhan@alibaba-inc.com}

\renewcommand{\shortauthors}{Chunqi Wang et al.}

\begin{abstract}
Discriminative recommendation tasks, such as CTR (click-through rate) and CVR (conversion rate) prediction, play critical roles in the ranking stage of large-scale industrial recommender systems.
However, training a discriminative model encounters a significant overfitting issue induced by data sparsity. 
Moreover, this overfitting issue worsens with larger models, causing them to underperform smaller ones.
To address the overfitting issue and enhance model scalability, we propose a framework named GPSD (\textbf{G}enerative \textbf{P}retraining for \textbf{S}calable \textbf{D}iscriminative Recommendation), drawing inspiration from generative training, which exhibits no evident signs of overfitting. 
GPSD leverages the parameters learned from a pretrained generative model to initialize a discriminative model, and subsequently applies a sparse parameter freezing strategy.
Extensive experiments conducted on both industrial-scale and publicly available datasets demonstrate the superior performance of GPSD.
Moreover, it delivers remarkable improvements in online A/B tests.
GPSD offers two primary advantages: 1) it substantially narrows the generalization gap in model training, resulting in better test performance; and 2) it leverages the scalability of Transformers, delivering consistent performance gains as models are scaled up.
Specifically, we observe consistent performance improvements as the model dense parameters scale from 13K to 0.3B, closely adhering to power laws.
These findings pave the way for unifying the architectures of recommendation models and language models, enabling the direct application of techniques well-established in large language models to recommendation models.
The code is available at \url{https://github.com/chqiwang/gpsd-rec}.
\end{abstract}



\maketitle



\section{Introduction}
\begin{figure}[!tbp]
    \centering
    \begin{subfigure}[b]{0.48\linewidth}
        \centering
        \includegraphics[width=\linewidth]{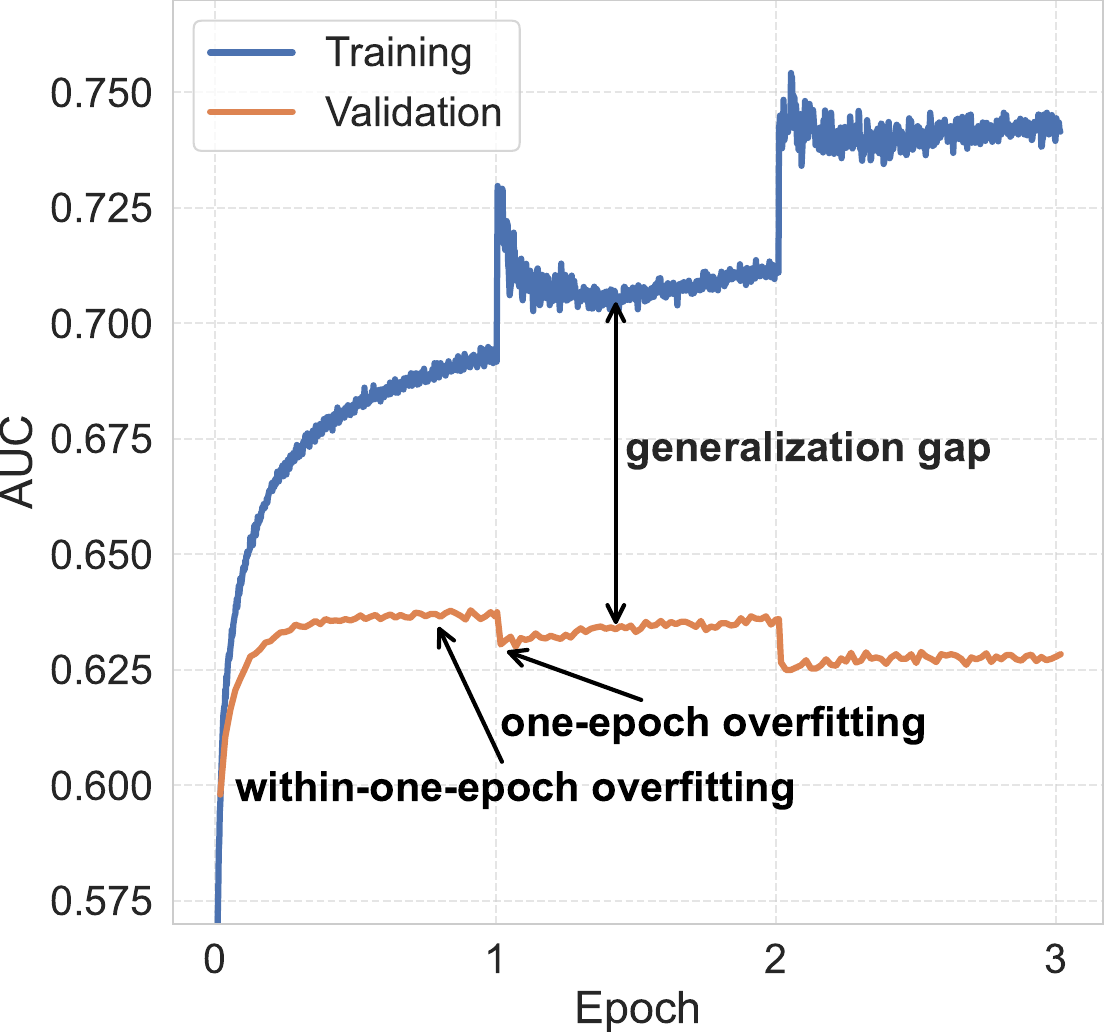}
        \caption{Performance during training.}
        \label{fig:intro_curve}
    \end{subfigure}
    \begin{subfigure}[b]{0.48\linewidth}
        \centering
        \includegraphics[width=\linewidth]{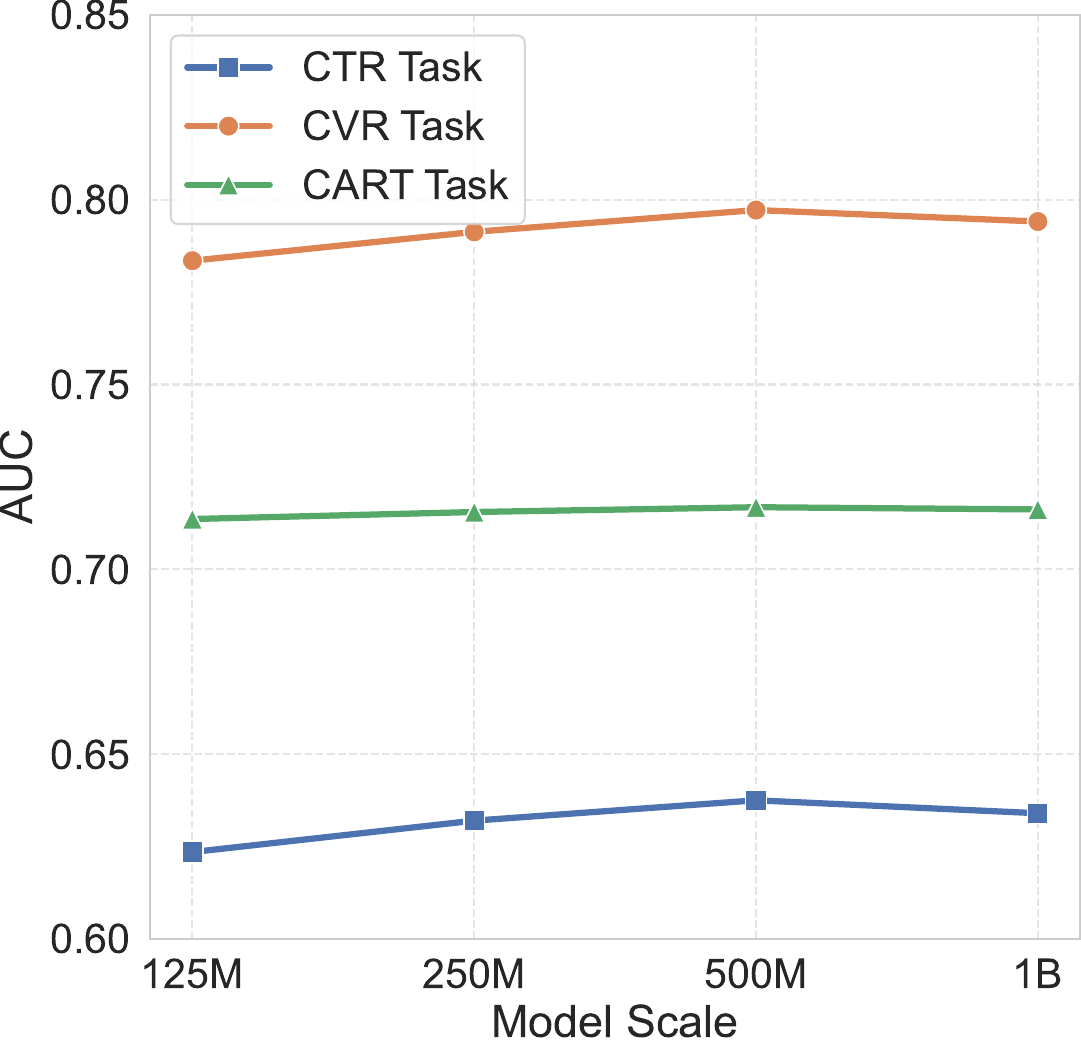}
        \caption{Performance across scales.}
        \label{fig:intro_scale}
    \end{subfigure}
    \caption{Illustration of the overfitting phenomenon (left) and the limited effectiveness of model scaling (right) for discriminative recommendation.}
    \Description{Illustration of the overfitting phenomenon (left) and the limited effectiveness of model scaling (right) for discriminative recommendation.}
    \label{fig:intro}
\end{figure}

Most industrial recommender systems follow a multistage pipeline, with the candidate retrieval and ranking phases being the most critical. The goal of the candidate retrieval phase is to retrieve a substantial number of items, ranging from ten to hundreds of thousands, from a vast item pool. In contrast, the ranking phase aims to select a limited set of items (dozens) that are most likely to interest the user from the candidates. The ranking model is typically discriminative and trained on items exposed to users, estimating engagement metrics such as CTR (click-through rate) and CVR (conversion rate). These metrics are then aggregated to determine the final recommendation list. These two phases correspond to two types of models: generative models and discriminative models.

\begin{figure*}[!htbp]
    \includegraphics[width=\textwidth]{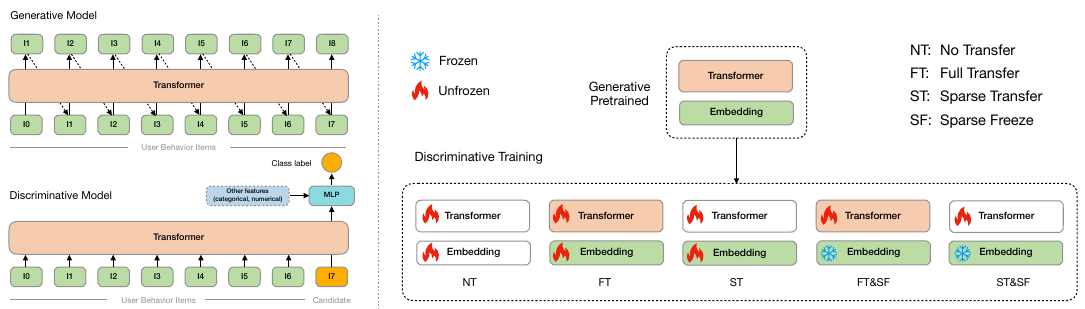}
    \centering
    \caption{Overview of the GPSD framework. The left part shows the generative and discriminative models based on the Transformer architecture and the right part illustrate the five strategies for bridging the generative pretraining and the discriminative training.}
    \Description{Overview of the GPSD framework.}
    \label{fig:framework}
\end{figure*}

In this paper, we focus on training discriminative models for recommendation. To obtain a superior discriminative model, a natural approach is to leverage powerful Transformer architecture \cite{vaswani2017attention} to encode user behavior items. The Transformer employs multiple stacked attention and feed-forward layers, significantly enhancing its modeling capacity, leading to remarkable successes in both language and vision domains \cite{vaswani2017attention,dosovitskiy2020image}. Moreover, the Transformer architecture has demonstrated strong scalability, with the discovery of scaling laws \cite{kaplan2020scaling} forming the foundation for the success of large language models \cite{dubey2024llama,yang2024qwen2,liu2024deepseek}.

However, training a Transformer-based discriminative recommendation model faces challenges. Although previous work \cite{chen2019behavior,gu2020deep,gu2021self} applied similar Transformer architecture to discriminative tasks, their model sizes were very small, with only single layer being used. None of them succeeded in leveraging the scalability of the Transformers.
By closely examining the metrics throughout the training process (refer to Figure \ref{fig:intro_curve}), we observe a significant generalization gap, which is a clear sign of overfitting. More specifically, there are two different types of overfitting phenomenons.
The first type is the sudden occurrence of overfitting at epoch transitions, referred to as \textit{one-epoch overfitting}. \cite{zhang2022towards} was the first to investigate this phenomenon and revealed that feature sparsity is the underlying cause.
The second type of overfitting is subtler, which starts at the early step within the first epoch and persists throughout the whole training process. Corresponding to one-epoch overfitting, we name this phenomenon \textit{within-one-epoch overfitting}.
The overfitting issue severely hinders the ability of Transformer-based discriminative models to achieve better performance by scaling the model size.
As illustrated in Figure \ref{fig:intro_scale}, there is a weak association between model scale and performance, which is in stark contrast to the scaling law observed in language models \cite{kaplan2020scaling}.
The limited effectiveness of model scaling has also been observed in \cite{ardalani2022understanding,guo2023embedding}.

Despite the severe overfitting challenge faced in training discriminative recommendation models, we observe that autoregressive generative models, that were trained to predict the next item conditioned on previous behavior items using sampled softmax loss \cite{jean2014using,wu2024effectiveness}, do not suffer from this problem. We hypothesize that generative training avoids the sparsity issue through extensive random negative sampling, thus leading to more stable and sufficient training of sparse parameters.
This discrepancy inspired us to propose a framework, named GPSD, which leverages generative pretraining to handle sparse parameters while focusing solely on dense parameters during discriminative training.
Our experiments demonstrate that the framework successfully addresses the overfitting issue and achieves substantial performance improvements across multiple industrial and public datasets, as well as in online A/B tests.
Furthermore, after addressing the overfitting issue, Transformer-based model performance improves consistently with increasing dense parameters from 13K to 0.3B, adhering to a predictable scaling law similar to that in language models \cite{kaplan2020scaling}.

The main contributions of this work are as follows:
\begin{itemize}
\item We revisit the overfitting phenomenon in recommendation models, showing two types of overfitting on an industrial-scale dataset. Additionally, we highlight the discrepancy in overfitting behavior between generative and discriminative models.
\item We propose a framework, named GPSD, which leverages generative pretraining and a strategy of freezing sparse parameters to effectively mitigate overfitting in discriminative models. GPSD achieves substantial performance improvements across multiple industry and public datasets, and it also achieves remarkable gains in online experiments.
\item We scale up the Transformer from 13K to 0.3B dense parameters for large-scale discriminative task and observe consistent performance improvements, establishing a scaling law for discriminative recommendation.
\end{itemize}

\section{Methodology}
In this section, we introduce the proposed GPSD framework based on the Transformer architecture. The framework consists of three parts: 1) the generative pretraining part, 2) the discriminative training part and 3) the bridging of the generative pretraining and discriminative training. Figure \ref{fig:framework} presents an overview of the framework.

\subsection{Generative Pretraining}
\label{sec:gpt}
Similar to GPT \cite{radford2018improving} in the language domain, during the generative pretraining stage, we train a Transformer model to generate user behavior item sequence autoregressively. To keep the description concise, we first introduce the case of item IDs, and the integration of more features will be discussed later.

\noindent\textbf{Task Description}
Given the dataset $D$, where each element is a chronological user behavior item sequence $X = \{x_1, x_2, \dots, x_L\}$ with length $L$, the objective of generative training is to minimize the negative log likelihood of $D$. The probability of each sequence is factorized using the chain rule. Therefore, the loss function for generative training is
$$\mathcal{L}=\sum_{X\in D}\sum_{l\in{1,\dots L}}-\log(p(x_l |x_{<l})),$$
where $p(x_l | x_{<l})$ is the probability of the next item conditioned on the previous items, given by the model.

\noindent\textbf{Model Architecture} Following recent work on large language models \cite{touvron2023llama}, we employ the Transformer \cite{vaswani2017attention} architecture and leverage various improvements that were subsequently proposed, including 1) Pre-Normalization \cite{xiong2020layer} for better training stability, 2) RMSNorm \cite{zhang2019root} for better performance, 3) RoPE \cite{su2024roformer} for extendable positional encoding and 4) SwiLU \cite{shazeer2020glu} as the activation function.

For generative training, we apply a causal mask to each attention operation, thereby rendering the Transformer unidirectional.
In addition to the generative approach, we can also adopt a denoising approach to train the network, similar to how BERT \cite{devlin2019bert} operates. In this case, a bidirectional Transformer will be used instead. By default, we use the generative approach, but we will compare both approaches by experiments.

\noindent\textbf{Model Training}
In training autoregressive models, the typical approach to modeling probabilities is to use the softmax function. However, in large-scale recommendation scenarios, it is impractical to do so because of the huge vocabulary. Therefore, we use the sampled softmax \cite{jean2014using,wu2024effectiveness} instead of the normal softmax, aiming to reduce computational and memory complexity. Formally, we replace the softmax based probability 
$$p(x_l | x_{<l})=\frac{\exp(f(x_l;x_{<l}))}{\sum_{v\in V}{\exp(f(v; x_{<l}))}}$$
with sampled softmax
$$\hat{p}(x_l | x_{<l})=\frac{\exp(f(x_l;x_{<l}))}{\exp(f(x_l; x_{<l}))+\sum_{n\in N}{\exp(f(n; x_{<l}))}},$$
where $V$ is the full set of items, $N$ is the negative sampled set of items, and $f(\cdot)$ represents the logit given by the models. We use a uniform sampler to sample negative items so the correction term is omitted.

To further reduce memory usage, we share the negative samples within each sequence and tie the embedding layer and the output linear layer. We use BFloat16 for training, which results in only slight loss compared to Float32 training while halving memory usage and accelerating training.
We train the model using the AdamW optimizer \cite{loshchilov2019adamw}. We use a linear warmup to reach a peak learning rate followed by cosine decay to 10\% of the peak.

\noindent\textbf{Integrating Side Features}
We have only consider item IDs as the model input by now. However, side features, such as category Ids, are also crucial in real-world recommender systems. To integrate these features, we make two adaptations to the model.
The first adaptation is on the embedding layer. Each feature is independently mapped to an embedding, and all embeddings are summed to form the Transformer input.
The second adaptation involves the loss component. In addition to the next item ID, we can also train the model to predict the next item's features. This results in multiple losses, which are then aggregated to form the final loss.

\subsection{Discriminative Training}
Discriminative models play a crucial role in the ranking phase of industrial recommender systems. We adopt a similar Transformer-based architecture for discriminative models as that used in generative models, with some minor modifications, which will be discussed in this section.

\noindent\textbf{Task Description}
Discriminative recommendation models take multiple features as input and output the probability over several classes. Input features can be divided into three groups: 1) the user behavior items, 2) the candidate item and 3) other categorical and numerical features.

\noindent\textbf{Model Architecture}
We concatenate the user behavior items with the candidate item to form the input sequence, which is then fed into the Transformer. To enable the Transformer to better distinguish between the user behavior items and the candidate item, we add an extra segment embedding onto the item embedding.
We also attach an MLP head on top of the last Transformer layer so that other categorical and numerical features can also be processed.

In this stage, we can also choose to use either a unidirectional Transformer or a bidirectional Transformer. We employ a unidirectional Transformer by default for better online inference efficiency and we will conduct experiments to compare the performance of both choices.

\noindent\textbf{Model Training}
We use cross-entropy as the loss function and other training settings remain the same as in the pretraining stage.

\subsection{Bridging Generative Pretraining and Discriminative Training}
\label{section:bridge}
In the language domain, It is widely recognized that pretraining a large Transformer model on a large unlabeled corpus and then simply transfering and finetuning all the parameters on task specific datasets achieves superior task performance \cite{devlin2019bert}. However, in the recommendation domain, this statement may not hold true and it is necessary to consider fine strategies.

As the sparse parameters, i.e., the embedding table, play a critical role in recommendation models and cause a lot of discrepancy between recommendation and language domain, we take meticulous care of it by splitting model parameters to sparse and dense parts.
We take the following five strategies when transfer a pretrained generative model to a discriminative model:
\begin{itemize}
    \item \textit{No Transfer (NT)}: Train all parameters from scratch. This serves as a baseline.
    \item \textit{Full Transfer (FT)}: Transfer all parameters, including both sparse and dense, from the pretrained generative model.
    \item \textit{Sparse Transfer (ST)}: Transfer the sparse parameters from the pretrained generative model while the dense parameters are trained from scratch.
    \item \textit{Full Transfer \& Sparse Freeze (FT\&SF)}: Apply the \textit{FT} strategy and freeze the sparse parameters during training.
    \item \textit{Sparse Transfer \& Sparse Freeze (ST\&SF)}: Apply the \textit{ST} strategy and freeze the sparse parameters during training.
\end{itemize}
See Figure \ref{fig:framework} for a better illustration of the five strategies.

\section{Experiments}
\subsection{Datasets}
We adopt both industrial datasets and publicly available datasets for experiments, which are listed in Table \ref{tab:datasets}.

\noindent\textbf{Industrial Datasets}
We consider three discriminative tasks, including click-through rate prediction, conversion rate prediction, and cart prediction, corresponding to CTR, CVR and CART in Table \ref{tab:datasets} respectively. To further explore the capabilities of large models, we collect CTR-XL, a larger dataset for click-through rate prediction, which contains 5B samples. 
Each dataset is partitioned temporally, with the data in most recent day allocated to the validation and test sets, while preceding data constitutes the training set.
For the generative pretraining of above tasks, we collect a separate dataset named CLICK. To construct CLICK, we first sort the clicked items of each user in chronological order, and then segment them into subsequences of a specific length range.

\begin{table}
    \caption{Datasets statistics. $L_{min}$ and $L_{max}$ are the minimum and maximum lengths of item sequences.}
    \label{tab:datasets}
    \scalebox{0.88}{
        \begin{tabular}{llccccc}
            \toprule
            Source&Datasets&\#Samples&\#Tokens&Vocab&$L_{min}$&$L_{max}$\\
            \midrule
            \multirow{4}{*}{Industry}&CLICK & 278M & 27B & 4M & 50 & 100\\
            &CTR & 1.6B & - & 4M & 10 & 100\\
            &CVR & 68M & - & 4M & 10 & 100\\
            &CART & 126M & - & 4M & 10 & 100\\
            &CTR-XL & 5B & - & 4M & 10 & 100\\
            \hline
            \multirow{3}{*}{Public}&Taobao & 969k & 39M & 4.1M & 10 & 200 \\
            &Electronics & 192k & 1.2M & 63k & 2 & 50 \\
            &Foods & 127k & 890K & 41k & 2 & 50 \\
            \bottomrule
        \end{tabular}
    }
\end{table}

\noindent\textbf{Public Datasets}
We choose three publicly available real-world datasets for experiments. (1) Taobao dataset contains user behaviors collected from Taobao\footnote{ https://tianchi.aliyun.com/dataset/dataDetail?dataId=649} in 9 days.
Following \cite{MIMN}, we chronologically organize clicked items to construct user behavior sequences for each user.
Assuming a user has clicked T items, we use the T-th clicked item as the positive label and randomly sample an item as the negative label. 
In addition, we regard the preceding T-1 items as the user behavior sequence, which is used to generate the pretraining dataset.
(2) Amazon dataset collects product reviews and metadata from Amazon\footnote{http://jmcauley.ucsd.edu/data/amazon/}.
We conduct experiments on the Electronics and Foods subsets, treating product reviews as user click sequences. 
The construction of these subsets aligns with that of the Taobao dataset.

\noindent\textbf{Feature Set}
In these datasets, each item (including the candidate items and those in behavior sequences) is associated with an item ID and several side features like category ID. Other than that, these datasets do not contain other categorical and numerical features.

\subsection{Settings}
\noindent\textbf{Model Specification}
Since we employ the standard Transformer, we can denote each model with a standard code. We use L\textbf{u}H\textbf{v}A\textbf{w} to denote a model with \textbf{u} the model depth, \textbf{v} the model width, \textbf{w} the number of attention heads.

\noindent\textbf{Hyperparameters}
We use the hyperparameters listed in table \ref{tab:hyperparams}. The following provides some explanations. For industrial experiments, we use batch sizes of 32K for CTR/CTR-XL training, 16K for pretraining, and 4K for CVR/CART training. For pretraining and CTR/CVR/CART training, the learning rate is set to 5e-4. For CTR-XL training, the learning rate varies based on the model size (details in Table \ref{tab:model_scales}). Training epochs is set to 5 for pretraining, 3 for CTR/CVR/CART training and 1 for CTR-XL training.
For public datasets, we perform grid search to optimize learning rates.

\begin{table}
    \caption{Hyperparameters used in the experiments.}
    \label{tab:hyperparams}
    \scalebox{0.9}{
        \begin{tabular}{lcc}
            \toprule
            Hyperparameter&Industrial&Public\\
            \midrule
            AdamW Beta&(0.9,0.98)&(0.9,0.98)\\
            Weight Decay&0.1&0.1\\
            Peak Learning Rate&5e-4/1e-4/5e-5&5e-3/5e-4/1e-4\\
            Max Gradient Norm&1.0&1.0\\
            Batch Size&32K/16K/4K&512\\
            \#Warmup Steps&5000&500\\
            \#Training Epochs&1/3/5&10\\
            \#Negative Samples&4K&4K\\
            \bottomrule
        \end{tabular}
    }
\end{table}

\noindent\textbf{Hardware}
All models are trained on single or multiple A100 GPUs.

\noindent\textbf{Evaluation Metric}
We use AUC as the metric to evaluate discriminative models, which is the area under the ROC curve and is widely used in recommendation domain. It is not sensitive to classification threshold and a larger value means a better result.

\subsection{Revisiting the Overfitting Phenomenon}
In this section, we conduct experiments to show that discriminative recommendation models exhibit severe overfitting, while generative models do not exhibit this issue.

\begin{figure}[]
    \centering
    \begin{subfigure}[b]{\linewidth}
        \centering
        \includegraphics[width=\linewidth]{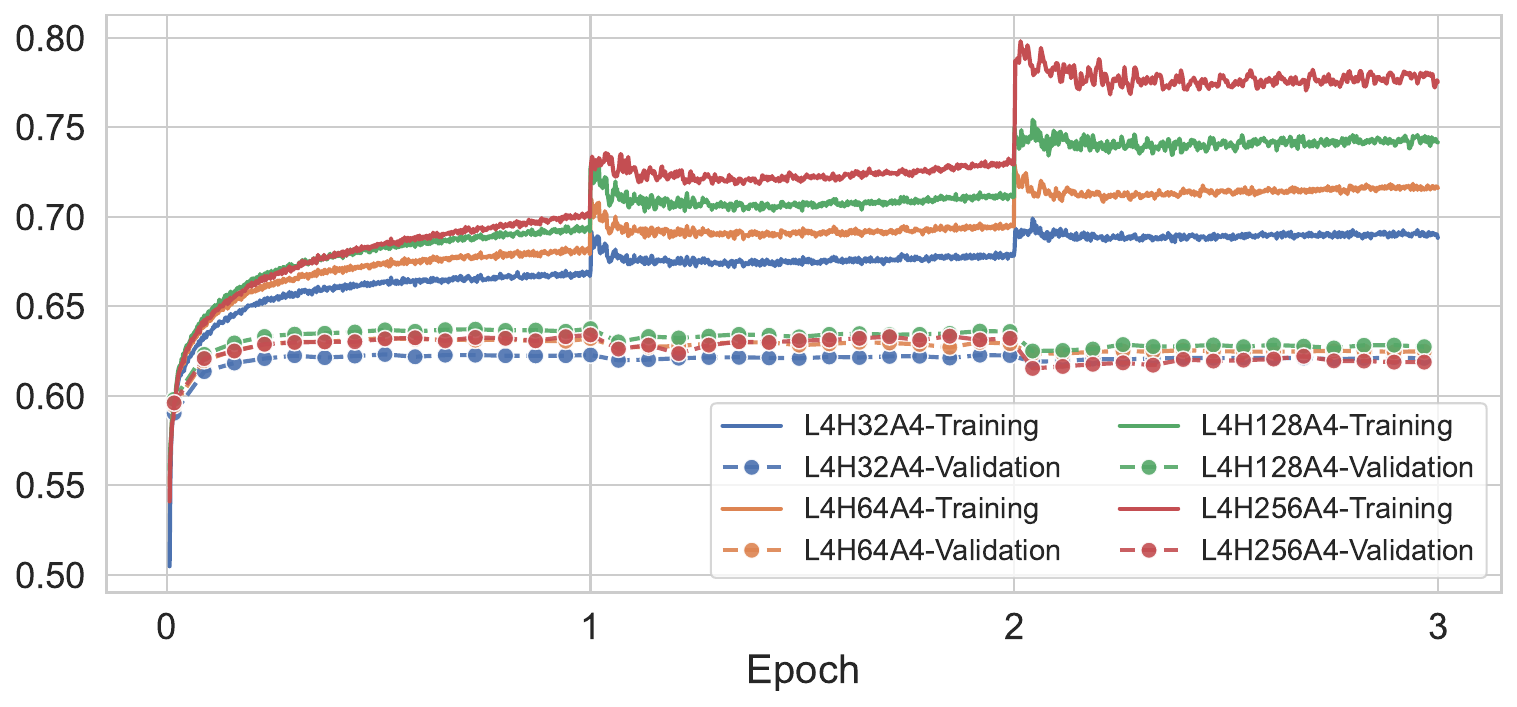} 
        \caption{AUC curves of discriminative models for the CTR task.}
        \label{fig:dis_gap}
    \end{subfigure}
    \begin{subfigure}[b]{\linewidth}
        \centering
        \includegraphics[width=\linewidth]{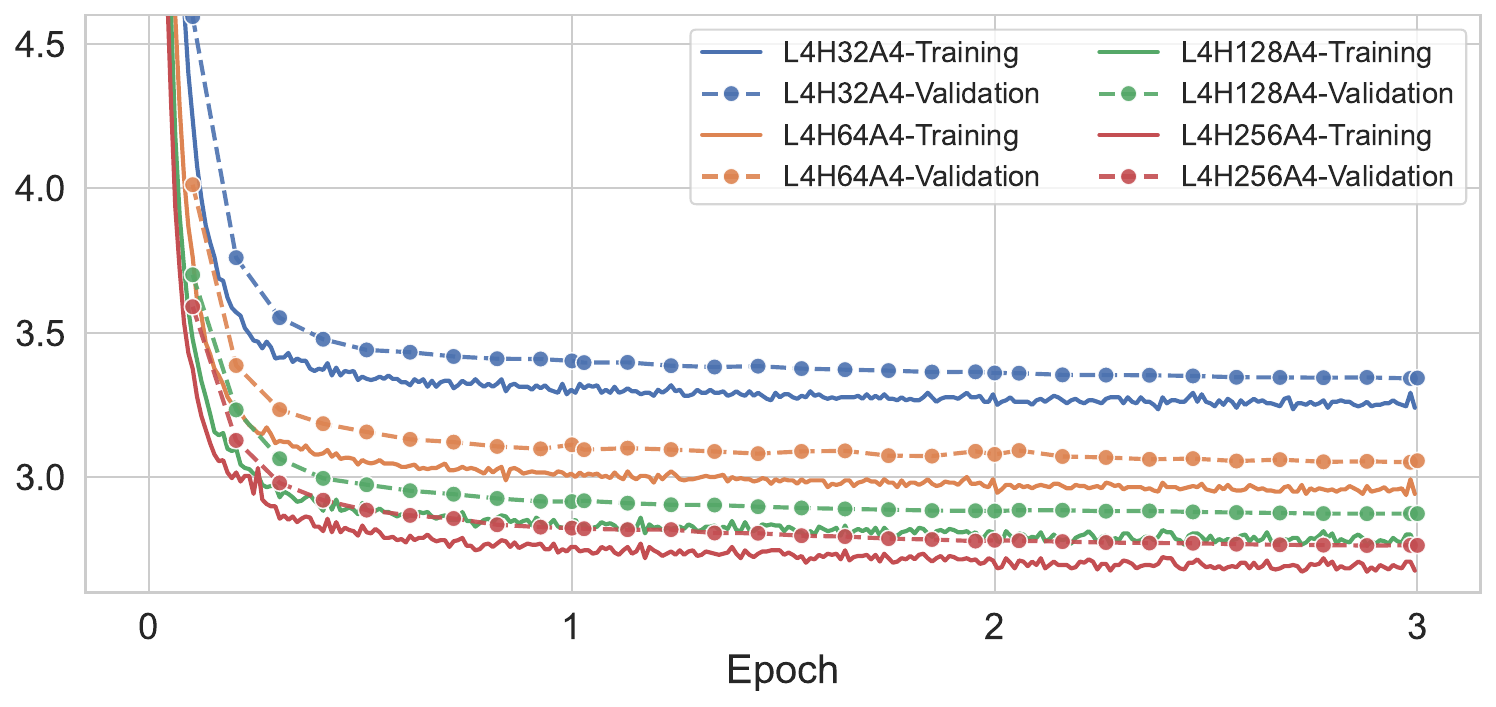}
        \caption{Loss curves of generative models.}
        \label{fig:gen_gap}
    \end{subfigure}
    \caption{A comparison of training behaviors between generative and discriminative models.}
    \Description{A comparison of training behaviors between generative and discriminative models.}
    \label{fig:overfitting}
\end{figure}

Figure \ref{fig:dis_gap} shows the training and validation AUC curves of discriminative models trained for the CTR task. Regardless of model scale, a significant generalization gap exists between training and validation performance, indicating severe overfitting. Consequently, while the larger model (L4H256A4) achieves significantly better training performance, it achieve worse validation performance than the smaller model (L4H128A4).
Specifically, we identify two distinct types of overfitting. As shown in Figure \ref{fig:dis_gap}, models across four scales all exhibit abrupt performance drop at epoch transitions. This phenomenon, referred to as \textit{one-epoch overfitting}, has been investigated in \cite{zhang2022towards}.
We confirm the occurrence of this phenomenon on a industrial scale dataset, which is $16\times$ the size of the largest dataset used in \cite{zhang2022towards}.
Besides \textit{one-epoch overfitting}, there is a second type of overfitting, which is subtler. After certain steps within the first epoch, the validation AUC almost stagnates while the training AUC continues to grow fast. We name this type of overfitting \textit{within-one-epoch overfitting} in order to correspond with \textit{one-epoch overfitting}.
Both overfitting phenomena hinder the scalability of Transformers for discriminative tasks and obstruct the path of replicating the successes of large language models achieved by scaling.

In contrast to discriminative models that face severe overfitting, we find generative models demonstrate robustness against this issue. As shown in Figure \ref{fig:gen_gap}, the training and validation loss curves of generative models remain a small constant gap throughout the training process. The constant generalization gap is expected and acceptable, often caused by distributional shifts occurring over time. This inherent resistance to overfitting leads to improved scalability, with larger models consistently achieve superior performance compared to their smaller counterparts.
We hypothesize that generative training avoids the sparsity issue through extensive random negative sampling, thus leading to more stable and sufficient training of sparse parameters.

\subsection{Enhancing Discriminative Training via Generative Pretraining}
\label{sec:enchancing}
Figure \ref{fig:overfitting} demonstrates that generative models, unlike discriminative models, do not face severe overfitting problem and can achieve better performance when scaled up.
This discrepancy inspires us to enhance discriminative training via generative pretraining.
As introduced in Section \ref{section:bridge}, there are various strategies to bridge a pretrained generative model and a discriminative model.
In this section, we conduct experiments on these strategies and try to figure out which one is better. The results are shown in Table \ref{tab:strategies}.
According to results, we can draw the following conclusions:
\begin{itemize}
    \item \textit{FT} and \textit{ST} strategies only lead to slightly better performance than training from scratch (\textit{NT}). This indicates that the pretraining and finetuning framework established in the language domain is not enough for recommendation tasks.
    \item Freezing sparse parameters (\textit{FT\&SF} and \textit{ST\&SF}) leads to significantly better performance than fully training (\textit{FT} and \textit{ST}) in most of the cases, indicating that sparse parameters learning is problematic in discriminative training.
    \item \textit{FT\&SF} and \textit{ST\&SF} cannot defeat one another in all scenarios. The results suggest that \textit{FT\&SF} can achieve better results either when the discriminative dataset is small or the model scale is large. In terms of flexibility, \textit{ST\&SF} offers significant advantages since it enables cross-architecture transfer and integration of incremental training, which will be presented in Sections \ref{sec:cross_transfer} and \ref{sec:online_results}.
    \item With \textit{FT\&SF} and \textit{ST\&SF} strategies, scaling the Transformer from L4H32A4 to L4H256A4 consistently results in better performance.
\end{itemize}

\begin{figure}[htbp]
    \centering
    \begin{subfigure}[b]{\linewidth}
        \centering
        \includegraphics[width=\linewidth]{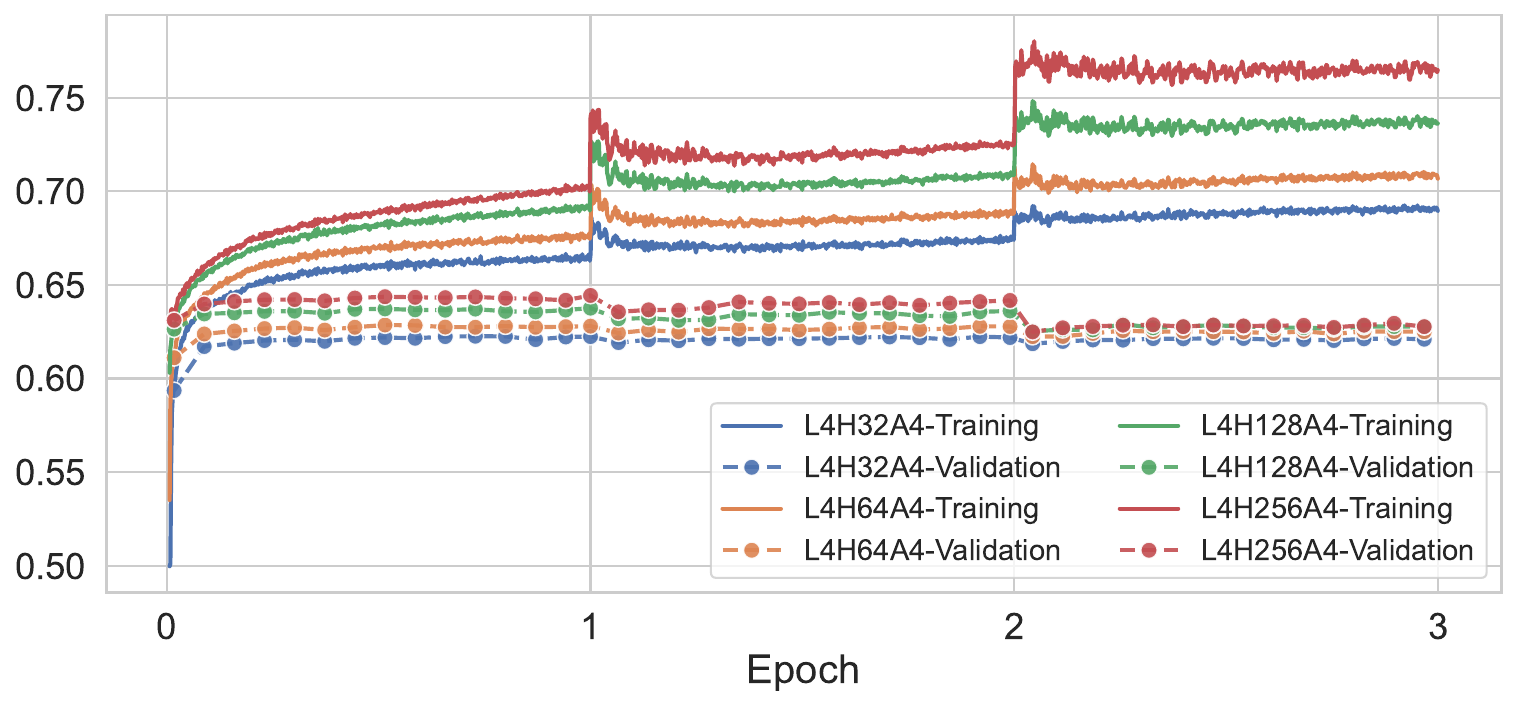}
        \caption{FT}
        \label{fig:generalization_gap_ft}
    \end{subfigure}
    \begin{subfigure}[b]{\linewidth}
        \centering
        \includegraphics[width=\linewidth]{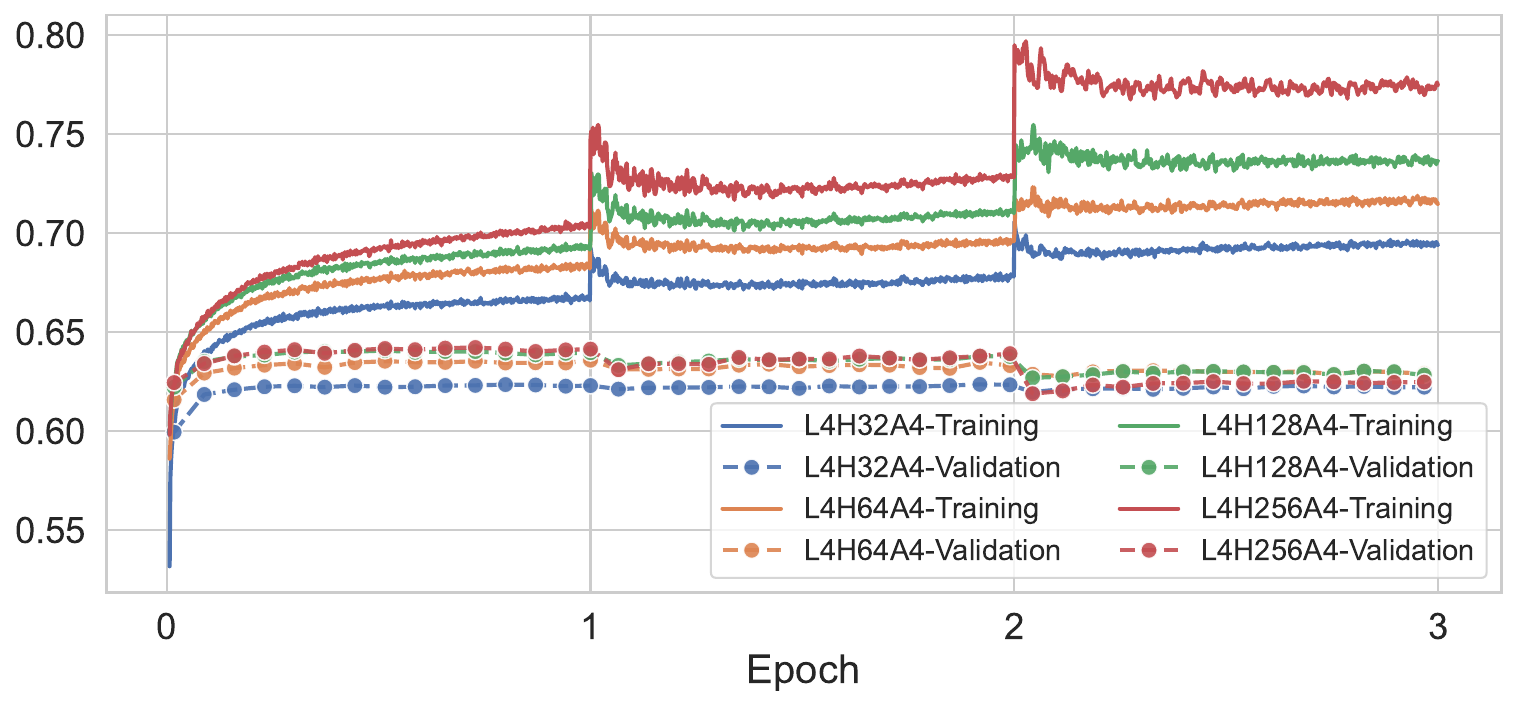}
        \caption{ST}
        \label{fig:generalization_gap_st}
    \end{subfigure}
    \begin{subfigure}[b]{\linewidth}
        \centering
        \includegraphics[width=\linewidth]{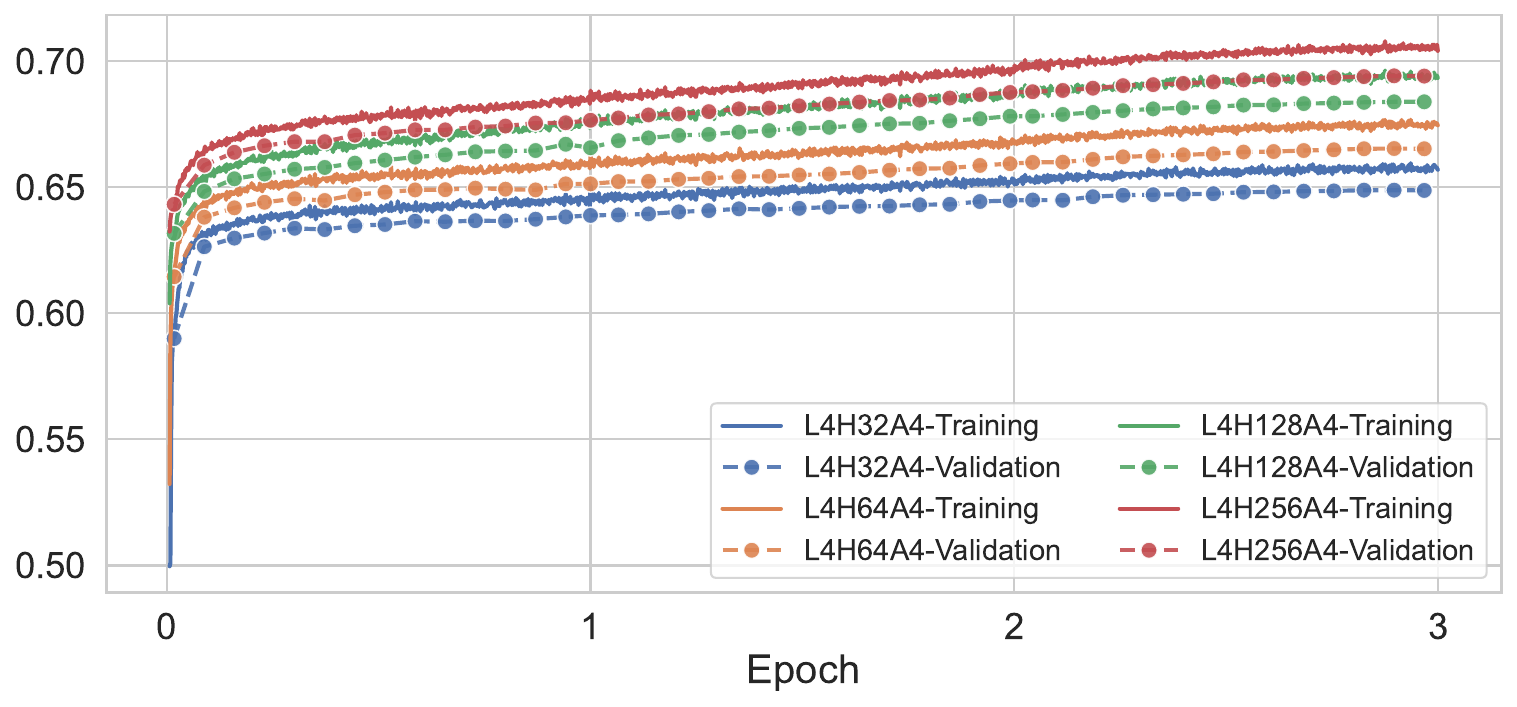}
        \caption{FT\&SF}
        \label{fig:generalization_gap_ftsf}
    \end{subfigure}
    \begin{subfigure}[b]{\linewidth}
        \centering
        \includegraphics[width=\linewidth]{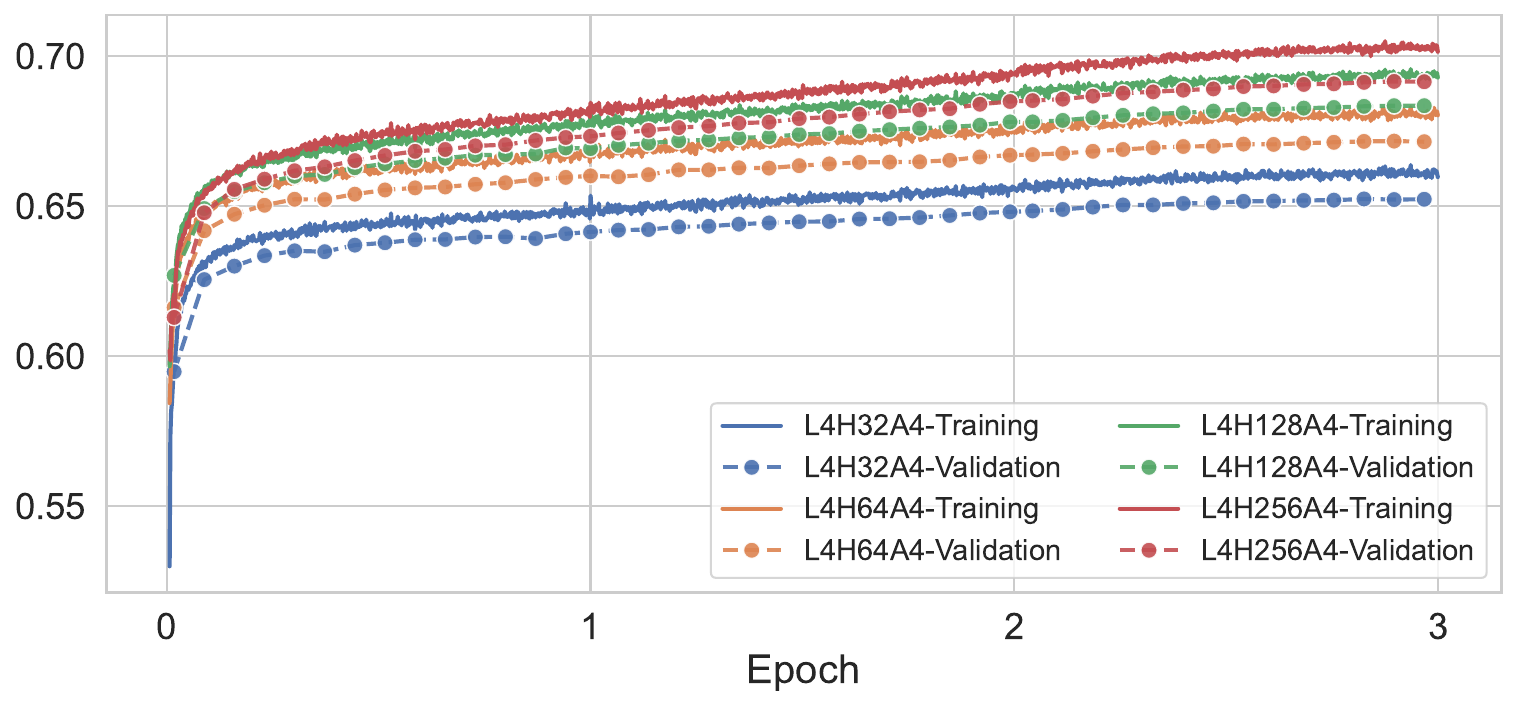}
        \caption{ST\&SF}
        \label{fig:generalization_gap_stsf}
    \end{subfigure}
    \caption{AUC curves of discriminative models for the CTR task under various strategies.}
    \label{fig:strategy_curve}
    \Description{AUC curves of discriminative models for the CTR task under various strategies.}
\end{figure}

To further demonstrate why the \textit{SF} strategy is effective, we present the AUC curves in Figure \ref{fig:strategy_curve}. Figures \ref{fig:generalization_gap_ft} and \ref{fig:generalization_gap_st} reveal that without the \textit{SF} strategy, the benefits of pretraining are limited, and the model suffers from the same overfitting issue as observed in the baseline model (Figure \ref{fig:dis_gap}). Figures \ref{fig:generalization_gap_ftsf} and \ref{fig:generalization_gap_stsf} reveal that the \textit{SF} strategy successfully addresses both one-epoch and within-one-epoch overfitting phenomena while substantially narrowing the generalization gap, leading to significantly better test performance.

\begin{table}
    \caption{Results on industrial datasets. The underline denote the best performance within each scale and bold denote the best performance across all scales. Imp represents the relative improvement.}
    \label{tab:strategies}
    \scalebox{0.8}{\begin{tabular}{llcccccc}
        \toprule
        \multirow{2}{*}{Model Scale}&\multirow{2}{*}{Strategy}&\multicolumn{2}{c}{CTR}&\multicolumn{2}{c}{CVR}&\multicolumn{2}{c}{CART}\\
        \cmidrule(lr){3-8}
        &&AUC & Imp & AUC & Imp & AUC & Imp\\
        \midrule
        \multirow{5}{*}{L4H32A4} & NT & 0.6235 & - & 0.7835 & - & 0.7136 & -\\
        &FT &   0.6231 & -0.06\% & 0.7887 & 0.66\% & \underline{0.7156} & 0.28\%\\
        &ST &   0.6238 & 0.05\% & 0.7857 & 0.28\% & 0.7152 & 0.22\%\\
        &FT\&SF &0.6487 & 4.04\% & 0.8092 & 3.28\% & 0.7151 & 0.21\%\\
        &ST\&SF&\underline{0.6524} & 4.63\% & \underline{0.8110} & 3.51\% & 0.7146 & 0.14\%\\
        \midrule
        \multirow{5}{*}{L4H64A4} & NT & 0.6320 & - & 0.7913 & - & 0.7155 & -\\
        &FT & 0.6286 & -0.54\% & 0.7932 & 0.24\% & 0.7185 & 0.42\%\\
        &ST & 0.6357 & 0.59\% & 0.7991 & 0.99\% & 0.7187 & 0.45\%\\
        &FT\&SF & 0.6654 & 5.28\% & 0.8204 & 3.68\% & 0.7274 & 1.66\%\\
        &ST\&SF & \underline{0.6716} & 6.27\% & \underline{0.8249} & 4.25\% & \underline{0.7281} & 1.76\%\\
        \midrule
        \multirow{5}{*}{L4H128A4} & NT & 0.6375 & - & 0.7972 & - & 0.7168 & -\\
        &FT & 0.6376 & 0.02\% & 0.8016 & 0.55\% & 0.7221 & 0.74\%\\
        &ST & 0.6395 & 0.31\% & 0.7997 & 0.31\% & 0.7212 & 0.61\%\\
        &FT\&SF & \underline{0.6838} & 7.26\% & 0.8290 & 3.99\% & \underline{0.7370} & 2.82\%\\
        &ST\&SF & 0.6834 & 7.20\% & \underline{0.8326} & 4.44\% & 0.7362 & 2.71\%\\
        \midrule
        \multirow{5}{*}{L4H256A4} & NT & 0.6340 & - & 0.7941 & - & 0.7162 & -\\
        &FT & 0.6444 & 1.64\% & 0.8050 & 1.37\% & 0.7244 & 1.14\%\\
        &ST & 0.6413 & 1.15\% & 0.8009 & 0.86\% & 0.7205 & 0.60\%\\
        &FT\&SF & \underline{\textbf{0.6940}} & 9.46\% & \underline{\textbf{0.8378}} & 5.50\% & \underline{\textbf{0.7437}} & 3.84\%\\
        &ST\&SF & 0.6916 & 9.09\% & 0.8327 & 4.86\% & 0.7381 & 3.06\%\\
        \bottomrule
    \end{tabular}}
\end{table}

\subsection{Comparison of Bidirectional and Unidirectional Transformers}
As mentioned in Section \ref{sec:gpt}, the generative approach, which uses unidirectional Transformers, is not the only method for pretraining. An alternative is the denoising approach, which employs bidirectional Transformers as introduced in BERT \cite{devlin2019bert}.
In addition, in the discriminative training stage, a bidirectional Transformer can also be adopted in place of the unidirectional one. To evaluate the impact of these alternatives, we conduct comparative experiments based on the L4H64A4 architecture.

\begin{table}
    \caption{Results of the unidirectional ($\rightarrow$) and bidirectional ($\rightleftarrows$) variations of L4H64A4 for the CTR task.}
    \label{tab:directional}
    \scalebox{0.95}{
        \begin{tabular}{lcccc}
            \toprule
            No.&Pretraining&Discriminative Training&Strategy&AUC\\
            \midrule
            (A)&  $\rightarrow$ & $\rightarrow$ & FT\&SF & 0.6654 \\
            (B)&  $\rightleftarrows$ & $\rightleftarrows$ & FT\&SF & 0.6662 \\
            (C)&  $\rightarrow$ & $\rightarrow$ & ST\&SF & 0.6716 \\
            (D)&  $\rightleftarrows$ & $\rightleftarrows$ & ST\&SF & 0.6666 \\
            (E)&  $\rightarrow$ & $\rightleftarrows$ & ST\&SF & 0.6707 \\
            (F)&  $\rightleftarrows$ & $\rightarrow$ & ST\&SF & 0.6673 \\
            \bottomrule
        \end{tabular}
    }
\end{table}

The results are presented in Table \ref{tab:directional}. A comparison between (A) and (B) reveals that bidirectional pretraining performs better than unidirectional pretraining when the FT\&SF strategy is employed, though the difference is marginal.
However, comparisons between (C) and (D) and between (E) and (F) indicate that unidirectional pretraining achieves superior performance to bidirectional pretraining when the \textit{ST\&SF} strategy is employed.
As for the discriminative training stage, comparisons between (C) and (E) and between (D) and (F) indicate that there is only a marginal performance gap between unidirectional and bidirectional Transformers.

Moreover, in practical industrial applications, unidirectional Transformers are often preferred due to their causal property, which facilitates significant efficiency improvements when combined with the KV cache technique \cite{pope2023efficiently}.
As a consequence, we adopt unidirectional Transformers as the default architecture for both pretraining and discriminative training.

\begin{table*}
    \caption{Settings for model scaling experiments.}
    \label{tab:model_scales}
    \begin{tabular}{llccccccc}
        \toprule
        Architecture&Pretraining Architecture&\#Layers&Hidden Dim& \#Att Heads&\#Sparse Params&\#Dense Params&Peak LR\\
        \midrule
        L1H32A4& L4H32A4&  1 & 32 & 4 & 125M & 13K & 5e-4\\
        L4H32A4& L4H32A4&  4 & 32 & 4 & 125M & 53K & 5e-4\\
        L4H64A4& L4H64A4&  4 & 64 & 4 & 250M & 213K & 5e-4\\
        L4H128A4& L4H128A4& 4 & 128 & 4 & 500M & 850K & 5e-4\\
        L4H256A4& L4H256A4& 4 & 256 & 4 & 1B & 3.4M & 5e-4\\
        L8H512A8& L4H512A8& 8 & 512 & 8 & 2B & 27M & 5e-4\\
        L12H768A12& L4H768A12& 12 & 768 & 12 & 3B & 92M & 1e-4\\
        L24H1024A16& L4H1024A16& 24 & 1024 & 16 & 4B & 327M & 5e-5\\
        \bottomrule
    \end{tabular}
\end{table*}

\subsection{Scaling Up Further}

As shown in Section \ref{sec:enchancing}, we have successfully addressed the overfitting problem in discriminative training and we have also observed that the performance consistently grows as we scale the model from L4H32A4 to L4H256A4. In this section, we scale the model further in order to explore the capabilities of very large Transformers. We prepare a larger CTR dataset with 5 billion samples, i.e., CTR-XL in Table \ref{tab:datasets}, for corresponding experiments and we train each model for one epoch due to resource constrain.

We leverage the \textit{ST\&SF} strategy in this experiment due to its flexibility of decoupling the architecture of generative pretraining and discriminative training.
The model settings used are listed in Table \ref{tab:model_scales}. 
For the pretraining phase, we use networks fixed at 4 layers with varying widths. This reduces resource consumption while maintaining comparable downstream performance.
For discriminative training, we apply a smaller peak learning rate for larger models for better training stability.
We scale the sparse parameters from 125M to 4B ($32\times$) and scale the dense parameters from 13K to 327M ($25K\times$).

\begin{figure*}[!htbp]
    \begin{subfigure}[b]{0.48\linewidth}
        \centering
        \includegraphics[width=\linewidth]{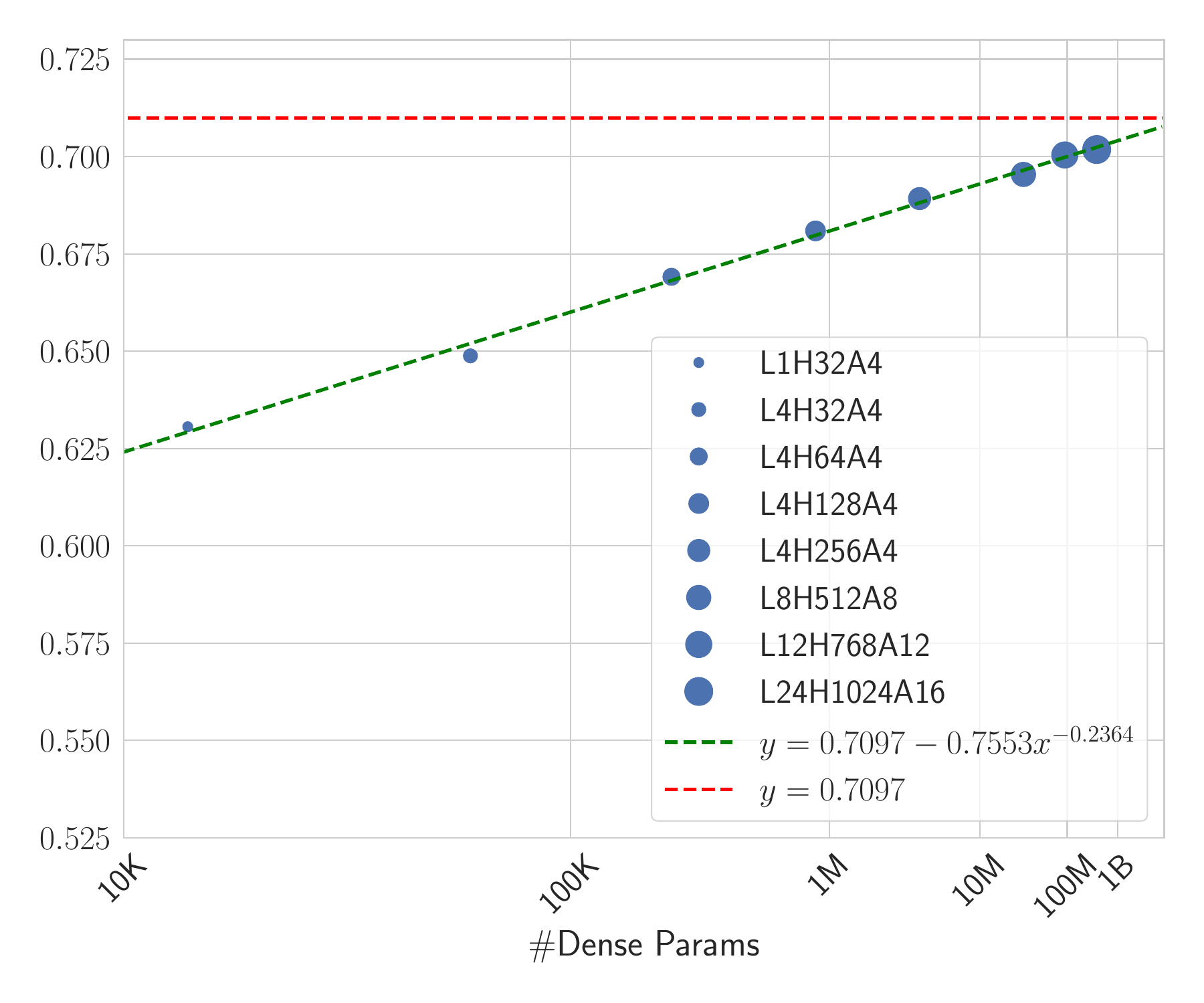}
        \caption{AUC}
    \end{subfigure}
    \begin{subfigure}[b]{0.48\linewidth}
        \centering
        \includegraphics[width=\linewidth]{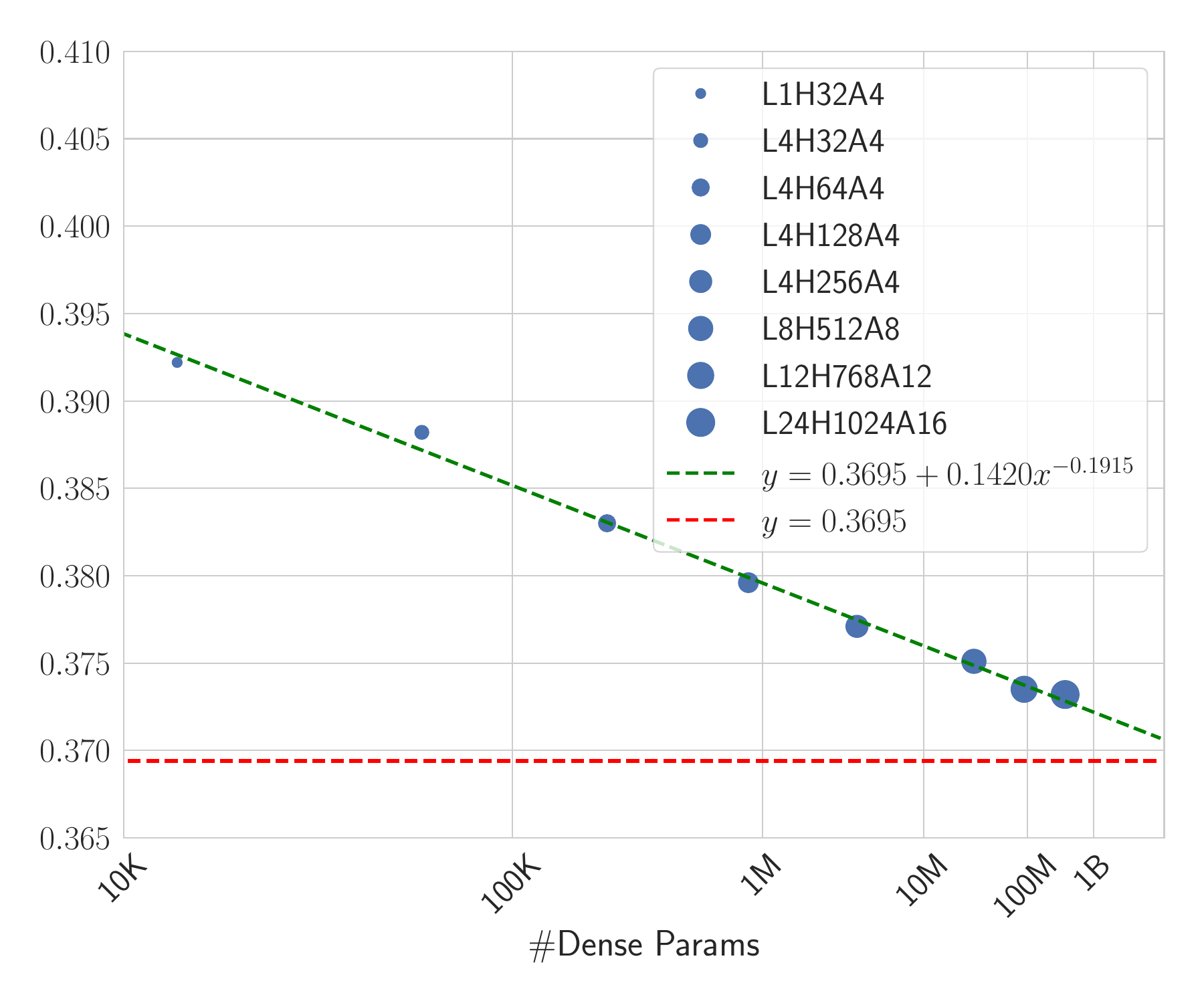}
        \caption{Loss}
    \end{subfigure}
    \caption{Results of scaling. Solid circles show the factual performance while the green dashed lines denotes the estimated power laws and the red lines are the derived empirical bounds.}
    \Description{Results of scaling.}
    \label{fig:scaling_law}
\end{figure*}

The results are shown in Figure \ref{fig:scaling_law}, which demonstrates the model performance consistently improves as we scale up the model size.
We also find that the power laws can fit the observations as shown by dashed lines in the figure.
The estimated power laws also tell us the empirical upper bound of AUC on the CTR-XL dataset is about 0.7097 and the empirical lower bound of loss is about 0.3695.
Detailed metrics and training costs are listed in Table \ref{tab:scaling_detail_metrics} of the appendix.

\subsection{Cross-Architecture Transfer}
\label{sec:cross_transfer}
\begin{figure}[htbp]
    \centering
    \begin{subfigure}[b]{0.48\linewidth}
        \centering
        \includegraphics[width=\linewidth]{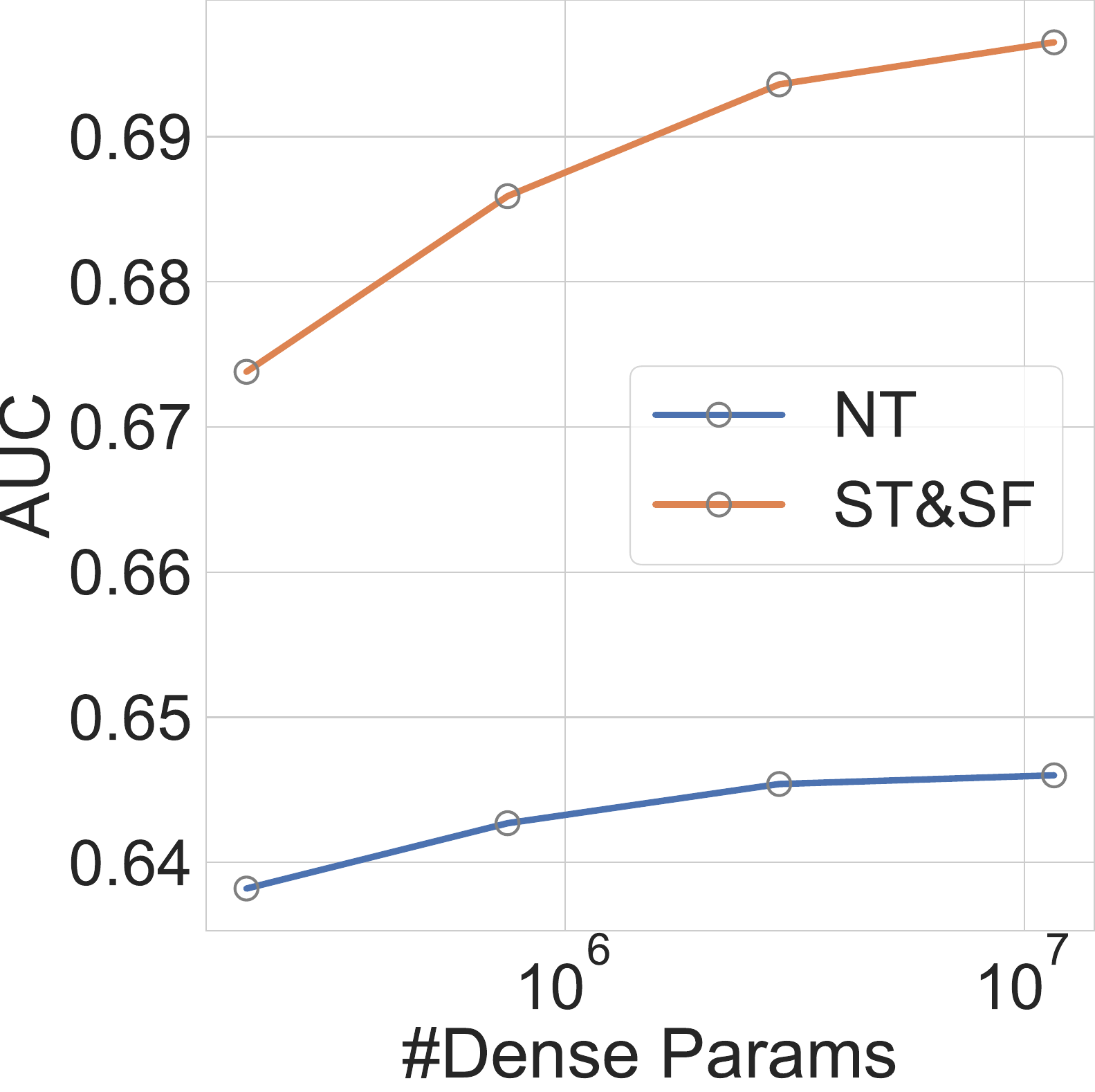}
        \caption{HSTU}
    \end{subfigure}
    \begin{subfigure}[b]{0.48\linewidth}
        \centering
        \includegraphics[width=\linewidth]{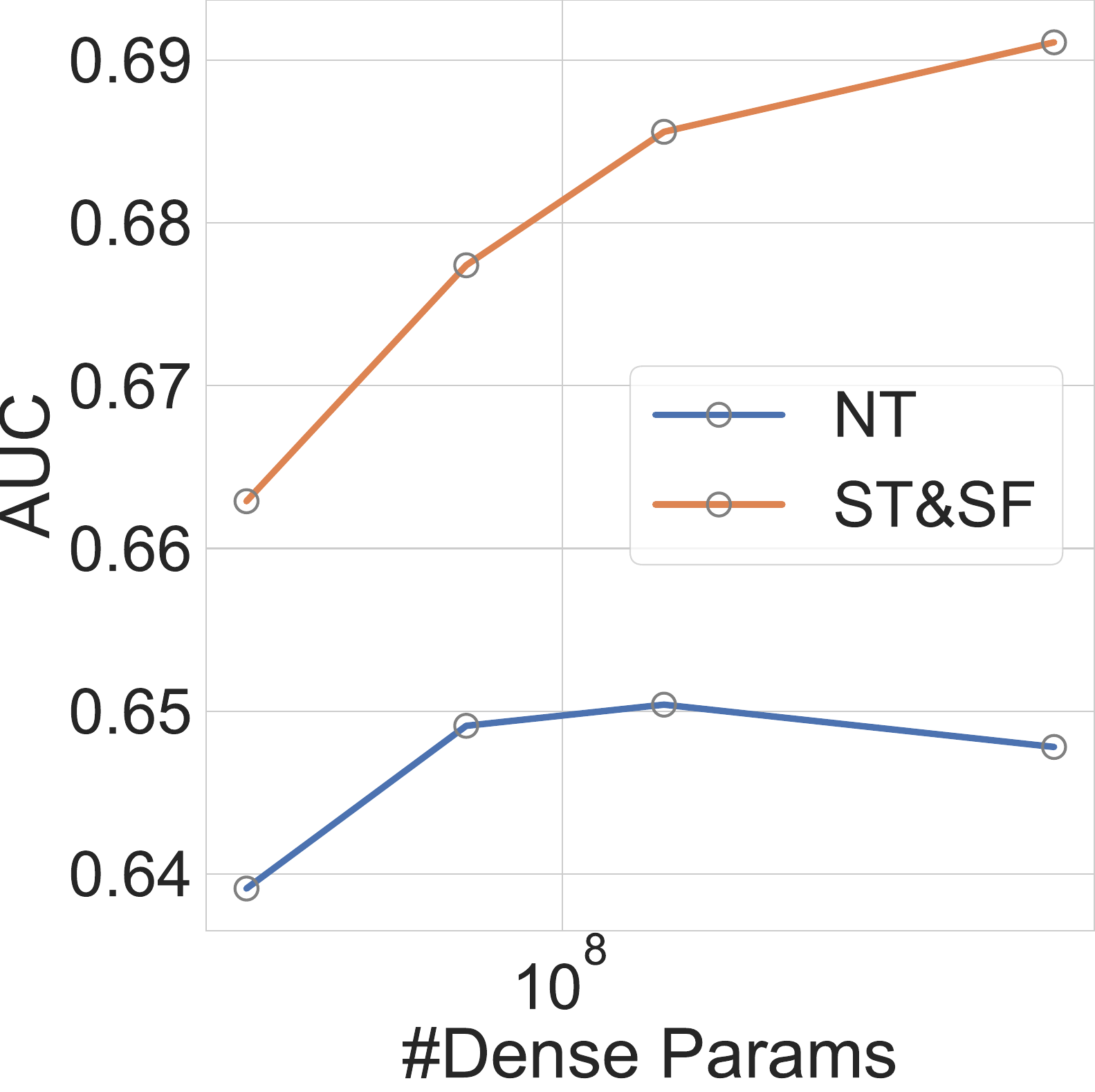}
        \caption{Wukong}
    \end{subfigure}
    \caption{Results of cross-architecture transfer from Transformer to HSTU and Wukong.}
    \Description{Results of cross-architecture transfer from Transformer to HSTU and Wukong.}
    \label{fig:cross_results}
\end{figure}

Considering the recent advancements in the recommendation community, some works have proposed new architectures that claim to successfully scale up.
we conduct experiments to transfer our pretrained sparse parameters in Transformer to these novel architectures by the $\textit{ST\&SF}$ strategy and analyze their impact on model performance and scalability.
We adopt two recent published architectures: HSTU \cite{zhai2024actions} and Wukong \cite{zhang2024wukong}.
HSTU is a sequential model capable of handling variable-length sequences, similar to Transformer, so we simply replace all Transformer layers with HSTU layers while keeping other components unchanged. However, Wukong is a non-sequential model. To meet its input format, we pad the input sequences to a fixed length and treat them as independent features.
For each the two architectures we have examined four different scales (refer to Table \ref{tab:cross_transfer} of the appendix for details).
The results are shown in Figure \ref{fig:cross_results}, from which we can see the \textit{ST\&SF} strategy substantially improves scalability of both HSTU and Wukong, despite the parameters are transfered from a different architecture, i.e., the Transformer.

\subsection{Results on Public Datasets}

\begin{table}
    \caption{Results on public datasets.
    Imp represents the relative improvement from combining the ST\&SF strategy.
    The bold denotes the best performance, and the underline denotes the greatest relative improvement.}
    \label{tab:public_result}
    \centering
    \scalebox{0.80}{
        \begin{tabular}{lcccccc}
            \toprule
            \multirow{2}{*}{Method} & \multicolumn{2}{c}{Taobao} & \multicolumn{2}{c}{Electronics} & \multicolumn{2}{c}{Foods}\\
            \cline{2-7}
            & AUC & Imp & AUC & Imp & AUC & Imp \\
            
            \midrule
            DeepFM          & 0.9044 & -        & 0.7985 & -        & 0.7425 &  -       \\ 
            DeepFM + ST\&SF & 0.9547 & 5.56\%   & 0.8665 & 8.52\%   & 0.7948 &  7.04\%  \\  \hline
            DIN             & 0.9341 & -        & 0.8091 & -        & 0.7711 &  -       \\
            DIN + ST\&SF    & 0.9752 & 4.40\%   & 0.8800 & 8.76\%   & 0.8223 &  6.64\%  \\ \hline
            DIEN            & 0.9386 & -        & 0.8055 & -        & 0.7803 &  -        \\
            DIEN + ST\&SF   & 0.9608 & 2.36\%   & 0.8788 & 9.09\%   & 0.8253 &  5.77\%  \\ \hline
            DMIN            & 0.9301 & -        & 0.7969  & -        & 0.7753 &  -        \\ 
            DMIN + ST\&SF   & 0.9620 & 3.43\%   & 0.8790 & 10.03\%  & 0.8317 &  7.27\%    \\  \hline
            DMR             & 0.9344 & -        & 0.8198 & -        & 0.7792 &  -       \\ 
            DMR + ST\&SF    & 0.9684 & 3.64\%   & 0.8779 & 7.08\%   & 0.8240 &  5.75\% \\  \hline
            L1H32A4             & 0.9310    & -             & 0.7964            &  -        & 0.7412            & -             \\
            L1H32A4 + ST\&SF    & 0.9708    & 4.27\%        & 0.8674            &  8.91\%   & 0.8228            & 11.01 \%      \\ \hline
            L4H32A4             & 0.9306    & -             & 0.7943            &  -        & 0.7508            & -             \\
            L4H32A4 + ST\&SF    & 0.9739    & 4.65\%        & 0.8715            &  9.72\%   & 0.8384            & 11.67\%       \\ \hline
            L4H64A4             & 0.9357    & -             & 0.7948            &  -        & 0.7495            & -             \\
            L4H64A4 + ST\&SF    & 0.9771    & 4.42\%        & 0.8810            &  10.84\%  & \textbf{0.8398}   & 12.05\%       \\ \hline
            L4H128A4            & 0.9257    & -             & 0.7944            &  -        & 0.7398            & -             \\
            L4H128A4 + ST\&SF   & 0.9806    & 5.93\%        & 0.8799            &  10.76\%  & 0.8368            & 13.11\%       \\ \hline
            L4H256A4            & 0.8651    & -             & 0.7710            &  -        & 0.7135            & -             \\
            L4H256A4 + ST\&SF   & \textbf{0.9808} & \underline{12.37\%} & \textbf{0.8847}  & \underline{14.47\%}& 0.8370  & \underline{17.31\%}\\
            \bottomrule
        \end{tabular}
    }
\end{table}
The overall perfromance on public datasets is shown in Table \ref{tab:public_result}.
We select some traditional models as baselines, including DeepFM \cite{deepfm}, DIN \cite{zhou2018deep}, DIEN \cite{DIEN}, DMIN \cite{dmin}, DMR \cite{dmr}. The embedding dimension for these baseline models are set to 64.

First, with the augment of the \textit{ST\&SF} strategy, all baseline models exhibit significant performance improvements, ranging from 2.36\% to 10.03\% across multiple datasets.
This demonstrates that our proposed framework exhibits strong compatibility with various recommendation models. Besides, due to the universality of the high-quality sparse parameters generated by the framework, it can be seamlessly deployed in a plug-and-play manner.

Furthermore, the results demonstrate that, without the \textit{ST\&SF} strategy, Transformer models perform notably worse than the baselines. However, when the proposed \textit{ST\&SF} strategy is integrated, Transformers achieve substantial improvements and surpass all baselines. This clearly validates the framework's effectiveness in mitigating overfitting.
Another noteworthy finding is that the largest Transformer model (L4H256A4) does not achieve the best performance across all datasets—likely due to the limited dataset scale—yet its performance remains competitive with the top-performing models. Most importantly, these large models no longer suffer from severe overfitting, further underscoring the robustness of our approach.

\subsection{Online A/B Results}
\label{sec:online_results}
\begin{figure}[!htbp]
    \includegraphics[width=0.45\textwidth]{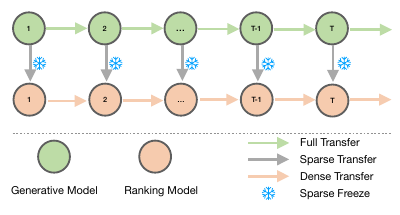}
    \centering
    \caption{Illustration of the incremental-GPSD framework. Both the generative model and the ranking model are trained in an incremental manner. In each iteration of the ranking model training, the sparse parameters from the generative model are first transferred to the ranking model and then frozen, while only the dense parameters are updated.}
    \Description{Illustration of the incremental-GPSD framework. Both the generative model and the ranking model are trained in an incremental manner. In each iteration of the ranking model training, the sparse parameters from the generative model are first transferred to the ranking model and then frozen, while only the dense parameters are updated.}
    \label{fig:igpsd}
\end{figure}

We apply the GPSD framework to the ranking model in the product recommender system of e-commerce platform AliExpress\footnote{https://www.aliexpress.com}. 
The base ranking model is a feature-rich multitask model, with hundreds of categorical and numerical features and three tasks. 
It employs single-layer target attention modules to encode user behavior items and is incrementally trained every day on new comming data. 
To apply the GPSD framework onto the base model, we develop an incremental-GPSD framework that integrates incremental training with the \textit{ST\&SF} strategy, as illustrated in Figure \ref{fig:igpsd}. 
We also substitute the target attention modules with Transformers to encode user behavior items. 
For online efficiency, we only employ small-scale Transformers, i.e., L3H160A4. 
Despite the small scale, the model still achieves significant online gains as shown in Table \ref{tab:online_result}.

\begin{table}[h!]
\centering
\caption{Online A/B test results based on a 30-day experimental window.}
\label{tab:online_result}
\begin{tabular}{l|cccc}
\toprule
Metric & GMV & Orders & Buyers & CTR \\
\midrule
Gains & +7.03\% & +2.11\% & +1.86\% & +3.78\% \\
\bottomrule
\end{tabular}
\end{table}

\section{Related Work}
\subsection{Sequential Recommendation}
Modeling user behavior sequence is the key to understand user interests for predicting potential prefered items, which is crucial for recommender systems.
In general, there are two types of tasks in this subject: retrieval tasks and ranking tasks.

Retrieval tasks target selecting a subset from a substantial collection of items that align with user preferences. 
A common approach is to train a model that can autoregressively to predict the next item, which is similar to autoregressive language modeling, rendering the model a generative model.
For example, GRU4Rec \cite{gru4rec} uses GRU based RNNs for next item prediction. Caser \cite{tang2018personalized} leverages CNNs for next item prediction. SASRec \cite{sasrec} employs unidirectional Transformers for next item prediction. In contrast, BERT4Rec \cite{bert4rec} leverages bidirectional Transformers that incorporate masked item prediction as an augmented task for next-item prediction.

Conversely, ranking tasks involve scoring candidate items with user behavior items. Ranking models are typically trained discriminatively on collected user action log, an example of which is CTR (click-through rate) prediction.
For example, DIN \cite{zhou2018deep} utilizes an target attention mechanism to capture the relationship between user behavior items and candidate items.
DIEN \cite{DIEN} further employs RNNs to capture temporal patterns in user behavior item sequences.
BST \cite{chen2019behavior} and DMT \cite{gu2020deep} leverages Transformers to model the user behavior item sequences.

More recently, \cite{zhai2024actions} propose a generative framework that unifies the architecture of retrieval and ranking models. However, in their framework, these models are trained independently.
In contrast, our framework bridges the training of retrieval and ranking models, addressing the scalability challenges in training ranking models.

\subsection{Overfitting in Recommendation Models}
Deep recommendation models based on the Embedding-MLP architecture are particularly susceptible to overfitting during training due to the presence of large sparse embeddings. The sparsity of user-item interactions can lead to models capturing noise rather than the underlying patterns, making the issue of overfitting critically important in recommender systems.

While numerous studies propose fancy architectures to enhance model performance, comparatively few have addressed the challenge of overfitting itself. \cite{zhang2022towards} highlights the intriguing one-epoch overfitting phenomenon observed in CTR models, demonstrating that commonly employed regularization techniques, such as dropout and weight decay, often fail to mitigate this problem effectively.
Subsequently, MEDA \cite{fan2024multi} successfully alleviates \textit{one-epoch overfitting} by reinitializing the embedding layer at the start of each training epoch.
In contrast to MEDA, we borrow advantages from generative training. In addition to \textit{one-epoch overfitting}, our proposed framework also addresses \textit{within-one-epoch overfitting}, narrowing the generalization gap to a small constant value during model training.

PeterRec \cite{yuan2020parameter} and SRP4CTR \cite{han2024enhancing} also improve performance in recommendation tasks through pretraining approach. However, they do not address the overfitting issue or aim to scale up their models to further enhance performance.

\subsection{Scaling Recommendation Models}
Recent studies \cite{kaplan2020scaling} have found that the performance of language models based on the Transformer architecture can steadily increase with the scaling of model size and data size, and it is even possible to predict the performance of larger models according to results of smaller models by power laws.
Besides language models, similar phenomenon have also been observed in vision models \cite{zhai2022scaling}.

However, in the realm of recommendation, parameter scaling seems not working well, particularly for discriminative tasks where severe overfitting issues occur. For example, BST \cite{chen2019behavior} uses Transformer to encode user sequence, reporting the best result on single-layer Transformer. This is also supported by DMT \cite{gu2020deep} and ZEUS \cite{gu2021self}, both of which also adopt single-layer Transformers. \cite{ardalani2022understanding} conclude that in recommendation domain, parameter scaling is running out of steam and does not contribute much to performance improvement.
\cite{guo2023embedding} also points out the model scalability issue in recommender systems and discovers the embedding collapse phenomenon that hurts model scalability.
Feature and data scaling is still the mainstream approach to enhance performance of recommendation models in industry \cite{mudigere2022software}, not parameter scaling.

Recently, new architectures tailored for recommendation tasks have been proposed \cite{zhai2024actions,zhang2024wukong}, leveraging ideas from the Transformer architecture to achieve better scalability. However, our experiments reveal that these architectures still face limitations when scaled to a certain extent. By incorporating generative pretraining, we can unlock significantly greater potential from these models.
Addtionally, \cite{zhang2024scaling,shin2023scaling} have also successfully scaled up recommendation models. However, they are either text-based models or specifically tailed for generation tasks.

\section{Conclusion and Future Work}
In this work, we address the critical challenge of overfitting in discriminative training for recommendation, which has long hindered the scalability of industrial recommendation models.
We propose a framework named GPSD. This framework leverages the parameters learned from a pretrained generative model to initialize a discriminative model, and applies a freezing sparse parameters strategy subsequently.
GPSD effectively mitigates the overfitting issue and brings remarkable performance and scalability improvement over Transformer-based models.
Extensive experiments show that GPSD achieves superior performance across multiple industrial and publicly available datasets, and obtains significant online gains.
Furthermore, by scaling the Transformer from 13K to 0.3B dense parameters, we observe steady performance improvements that adhere to power laws. These results bridge the gap between the architectures of recommendation and language models. Based on this work, techniques well-established in large language models can be directly applied to recommendation models.

This work has several limitations. First, we adopt a relatively small sequence length in our experiments. Second, we did not examine how the backbone model would affect the performance. For example, one can substitute the Transformer model with HSTU since the GPSD framework is model-agnostic and can be applied to any sequential model. Finally, due to efficiency issue, we have not yet deployed a very large-scale model online. We look forward to advanced engineering optimizations and thus enabling deploying larger models.

In the future, we would like to introduce advanced techniques established in training large language models into the training of recommendation models while further scaling up model size and sequence length.
We would also like to study how to transfer weights of state-of-the-art open-weights language models, such as Llama \cite{dubey2024llama}, Qwen \cite{yang2024qwen2} and Deepseek \cite{liu2024deepseek}, into ID based recommendation models.

\begin{acks}
We thank Hao Zhang, Jiahao Wang, and Zeji Zhou for their efforts in the online deployment.
\end{acks}
\bibliographystyle{ACM-Reference-Format}
\bibliography{gpsd}
\balance
\appendix
\newpage
\onecolumn
\section{Additional Materials}
\begin{table*}[!htbp]
    \caption{The detailed metrics and costs corresponding to Figure \ref{fig:scaling_law}.}
    \label{tab:scaling_detail_metrics}
    \begin{tabular}{lcccccc}
        \toprule
        ID&AUC$\uparrow$&Loss$\downarrow$&\#Training GPUs&\#Training Hours&\#Pretraining GPUs&\#Pretraining Hours\\
        \midrule
        L1H32A4&    0.6306&0.3922&4&3&2&14\\
        L4H32A4&    0.6488&0.3882&4&7&2&14\\
        L4H64A4&    0.6691&0.3830&4&10&4&17\\
        L4H128A4&   0.6809&0.3796&4&16&4&22\\
        L4H256A4&   0.6892&0.3771&4&29&4&39\\
        L8H512A8&   0.6954&0.3751&16&50&8&55\\
        L12H768A12& 0.7004&0.3735&16&101&8&82\\
        L24H1024A16& \textbf{0.7018}&\textbf{0.3732}&32&184&16&91\\
        \bottomrule
    \end{tabular}
\end{table*}

\begin{table*}[!htbp]
    \caption{Network settings for cross transfer.}
    \label{tab:cross_transfer}
    \begin{tabular}{l|ccc|ccccc}
        \toprule
        \multirow{2}{*}{ID}&\multicolumn{3}{c}{HSTU}&\multicolumn{5}{c}{Wukong}\\
        \cmidrule(lr){2-9}
        &\#Layers&Hidden Dim&\#Att Heads&\#Layers&Emb Dim&\#LCB Emb&\#FMB Emb&FMB Rank\\
        \midrule
        (A)& 8&64&4     &2&32&8&8&24\\
        (B)& 8&128&4    &4&64&32&32&24\\
        (C)& 8&256&4    &8&128&32&32&24\\
        (D)& 8&512&8    &8&256&48&48&48\\
        \bottomrule
    \end{tabular}
\end{table*}

\end{document}